\documentclass[reprint,aps,pre,groupedaddress,amsmath]{revtex4-2}

\usepackage{graphicx}% Include figure files
\usepackage{dcolumn}% Align table columns on decimal point
\usepackage{bm}% bold math

\begin{document}

\preprint{}

\title{Frank Elastic Anisotropy Induced by Anisotropic Attractive Interactions in Nematic Liquid Crystals}


\author{Akihiko Matsuyama}
\email[]{aki\_mcyama@mac.com}
%\thanks{}
\affiliation{Department of Physics and Information Technology,
Faculty of Computer Science and Systems Engineering, 
Kyushu Institute of Technology, 
Kawazu 680-4, Iizuka, Fukuoka 820-8502, JAPAN}


\date{\today}

\begin{abstract}

We present a mean-field theory to calculate the Frank elastic constants of a nematic liquid crystal with an anisotropic Maier--Saupe (MS) attractive interaction restricted to a cylindrical interaction volume of height $L_z$ and diameter $L_c$.
The shape of the interaction volume is characterized by the aspect ratio $\alpha_z=L_z/L_c$, which controls the spatial anisotropy of the attractive interaction. We derive expressions for the splay $(K_{11})$, twist $(K_{22})$, and bend $(K_{33})$ elastic constants and calculate them numerically as functions of concentration and temperature. At the special aspect ratio $\alpha_z=\sqrt{3}/2$, the transverse and axial second moments are equal, yielding the one-constant relation $K_{11}=K_{22}=K_{33}$ over the full ranges of concentration and temperature considered. Away from this point, oblate and prolate interaction volumes produce distinct elastic anisotropies. Changes in concentration or temperature can also reverse the hierarchy of the three elastic constants. 

\end{abstract}

%\keywords{}

\maketitle

\section{Introduction}
The Frank elastic constants are fundamental material parameters that characterize the free-energy costs of splay, twist, and bend deformations of the nematic director \cite{Oseen33,Frank58,Book_de, Ray10,MatsuCRC}. They govern a wide range of physical phenomena, including director fluctuations, defect structures, electro-optic responses, and hydrodynamic behavior in nematic liquid crystals \cite{Book_de}. Since Frank's pioneering continuum theory, considerable theoretical effort has been devoted to understanding the microscopic origins of these elastic constants \cite{Nehring72,Straley73,Priest73,Gruler75,Gruler72,Stecki80,Kimura81,Gelbart82,Meer82,Lee85,LeeMeyer88,Odijk86,Singh96,Capar13,Milchev18,Ferrarini18,Revignas24,Evangelista98,Marrucci91}.
For recent reviews of theoretical approaches to the Frank elastic constants, see Refs.~\cite{Varytimiadou24,Ferrarini10}.

For systems of hard rodlike molecules, Onsager established a molecular theory of nematic ordering based on excluded-volume interactions \cite{Onsager1949}. Building on this framework, Straley derived microscopic expressions for the Frank elastic constants of hard-rod nematics, demonstrating that elastic anisotropy originates from steric interactions between anisotropic particles \cite{Straley73,Poniewierski79,somoza89}. 
In contrast, many thermotropic liquid crystals are stabilized primarily by attractive intermolecular interactions, which are described by the Maier--Saupe (MS) theory \cite{maier59,Kyu95}.
The MS model has also been extended to calculate the Frank elastic constants of nematic phases \cite{Marrucci91,Varytimiadou24,Ferrarini10}.

For hard-rod nematics, Straley's theory based on excluded-volume interactions predicts the hierarchy \cite{Straley73,Lee85,LeeMeyer88}:
$
K_{33}>K_{11}>K_{22}. 
$
This hierarchy is observed in thermotropic nematic liquid crystals~\cite{HAKEMI83,bogi01,Bunning81}.
However, experimental and theoretical studies have revealed that this hierarchy is not universal.
Associating chromonic liquid crystals \cite{Nakata07,Zhou12,Zhou14,Zhou14b,Emanuele18,Matsu26},
polymeric lyotropic nematics \cite{Taratuta85,Taratuta88,Hakemi90,Sato96,Lucchetti16,Dietricha20}, and bent-core liquid crystals \cite{Kundu07,Sathyanarayana10,Majumdar11,Kaur12,Kaur13,yun15,Srigengan18,Iadlovska20,Babakhanova17,Cukrov17,Parsouzi19}
often exhibit different elastic anisotropies \cite{Warmerdam88,Czechowski04,Faetti09,Aya14,Oswald21,Louvergneaux26}.
These deviations indicate that, in addition to excluded-volume interactions, other molecular factors, such as attractive intermolecular interactions, molecular conformational flexibility, and intrinsic molecular curvature, play important roles in determining the elastic anisotropy~\cite{Evangelista98}.
For example, Cestari \textit{et al.} employed an MS molecular-field theory combined with atomistic molecular modeling to investigate the curvature elasticity of nematic liquid crystals \cite{Cestari09}.
However, the effects of anisotropic Maier--Saupe attractive interactions on the Frank elastic constants have not been systematically explored within a molecular-field framework.

In this paper, we develop a generalized MS model in which the attractive interaction is restricted to a cylindrical volume of height $L_z$ and diameter $L_c$. The aspect ratio $\alpha_z=L_z/L_c$ characterizes the shape of the interaction volume. By extending Straley's elastic theory \cite{Straley73} to this anisotropic interaction, we derive microscopic expressions for the Frank elastic constants and investigate how their magnitudes and relative ordering depend on concentration, temperature, and interaction-volume geometry. 
We find that the anisotropic attractive interactions provide a molecular mechanism for controlling both the magnitudes and the hierarchy of the Frank elastic constants.

Our model is motivated by the observation that attractive interactions between rod-like molecules can be anisotropic owing to their molecular architecture. For example, in liquid-crystalline molecules containing aromatic rings, the attractive interaction between face-to-face-stacked benzene rings differs from that between laterally adjacent rings. Similarly, in carbon nanotubes, the effective van der Waals interaction depends on the relative arrangement of the tubes. Our model captures these features through different interaction ranges within the plane spanned by the molecular orientations and along its normal direction, providing a simple description of anisotropic attractive interactions between rod-like molecules. Such anisotropic interaction ranges are physically well motivated because intermolecular forces generally depend on molecular geometry and relative orientation \cite{Israelachvili2011}.

In Section~\ref{sec2}, we introduce the Frank elastic constants for MS attractive interactions. In Section~\ref{sec3}, we present our numerical results for the elastic constants as functions of concentration and temperature. 
Section~\ref{sec4} summarizes our findings. 
In Appendix~\ref{app:strong-alignment}, we examine the geometrical factors under the strong-alignment approximation to elucidate the origin of the elastic anisotropy.

%%%%%%%%%%%%%Frank Elastic Constants of Nematic Liquid Crystals with Transverse Maier--Saupe Attractions
\section{Elastic constants with Maier--Saupe Attractions}
\label{sec2}
%%%%%%%%%%%%%%%%%%%%%%%%%%%%%%%%%%%%%%%%%%%%%%%%%%%%%%%%%%%%%
We consider a solution of identical rod-like molecules of length $L$ and diameter $D$ with the axial ratio $n=L/D$.
Let $f({\bf \Omega}\cdot{\bf n}({\bf r}))$ denote the orientational distribution function of the rod-like molecules, where ${\bf n}({\bf r})$ is the nematic director and ${\bf\Omega}$ is the orientation vector of a molecule located at position ${\bf r}$.
We describe the orientational free energy using a second-virial-type functional \cite{Onsager1949,MatsuCRC}, in which the pair interaction is represented by the Maier--Saupe attractive potential.
 The nematic free energy is given by
\begin{align}
\beta F
={}&
\rho \iint d{\bf r}\,d\Omega\,
f({\bf \Omega}\cdot{\bf n}({\bf r}))
\ln\!\left[4\pi f({\bf \Omega}\cdot{\bf n}({\bf r}))\right]
\nonumber\\
&+\frac{1}{2}\rho^2
\iiiint d{\bf r}\,d{\bf r}'\,d\Omega\,d\Omega'\,
f({\bf \Omega}\cdot{\bf n}({\bf r}))
f({\bf \Omega}'\cdot{\bf n}({\bf r}'))
\nonumber\\
&\qquad\times
U_{\rm MS}({\bf r},{\bf r}',{\bf \Omega},{\bf \Omega}').
\label{A1}
\end{align} 
where $\beta=1/(k_BT)$: $k_B$ is the Boltzmann constant and $T$ is the absolute temperature, $\rho=N/V$ is the number density of the rod, and $d\Omega=\sin\theta d\theta d\varphi$ is the solid angle, where $0<\theta<\pi$ and $0<\varphi<2\pi$. 
 The first term in Eq.~(\ref{A1}) shows the orientational entropy of the rod-like molecules and the second term represents the MS attractive interaction between two rod-like molecules located at ${\bf r}$ and ${\bf r}'$ with orientations ${\bf\Omega}$ and ${\bf\Omega}'$, respectively.

The excess free energy arising from spatial variations of the nematic director ${\bf n}({\bf r})$ is given by \cite{Straley73}
 \begin{eqnarray}
\beta \Delta F &= &
\frac{1}{2}\rho^2 \iiiint f({\bf \Omega}\cdot {\bf n}({\bf r}))   U_{MS}({\bf r}, {\bf r}', {\bf\Omega}, {\bf \Omega}')   
 \nonumber \\
 &&
 \times \big[ f({\bf \Omega}'\cdot {\bf n}({\bf r}')) -  f({\bf \Omega}'\cdot {\bf n}({\bf r})) \big]   d{\bf r} d{\bf r}' d\Omega d\Omega'.
 \label{A2}
 \end{eqnarray}
To obtain the Frank elastic free energy, Eq.~(\ref{A2}) is expanded to second order in the spatial derivatives of the director field\cite{Straley73}:
 %\begin{widetext}
\begin{equation}
\begin{aligned}
&f({\bf \Omega}'\cdot{\bf n}({\bf r}'))
-f({\bf \Omega}'\cdot{\bf n}({\bf r}))
\\
&= f'({\bf \Omega}'\cdot{\bf n}({\bf r}))
\bigg[
({\bf u}\cdot\nabla)({\bf \Omega}'\cdot{\bf n}({\bf r}))
\\
&\qquad
+\frac{1}{2}({\bf u}\cdot\nabla)^2
({\bf \Omega}'\cdot{\bf n}({\bf r}))
\bigg]
\\
&\quad
+\frac{1}{2}f''({\bf \Omega}'\cdot{\bf n}({\bf r}))
\big[
({\bf u}\cdot\nabla)({\bf \Omega}'\cdot{\bf n}({\bf r}))
\big]^2
+\cdots .
\end{aligned}
\label{A3}
\end{equation}
%\end{widetext}
where ${\bf u}={\bf r}'-{\bf r}$. The functions $f'(\cdots)$ and $f''(\cdots)$ denote the first and second derivatives of the orientational distribution function with respect to their argument.

 Substituting Eq. (\ref{A3}) into the free energy  Eq. (\ref{A2}), we obtain
% \begin{widetext}
\begin{equation}
\begin{aligned}
\beta \Delta F
&= \frac{1}{2}\rho^2 \iiiint
f({\bf \Omega}\cdot{\bf n}({\bf r}))
U_{MS}({\bf u};{\bf \Omega},{\bf \Omega}')
\\
&\quad \times
f'({\bf \Omega}'\cdot{\bf n}({\bf r}))
\big[({\bf u}\cdot\nabla)
({\bf \Omega}'\cdot{\bf n}({\bf r}))\big]
\,d{\bf r}\,d{\bf u}\,d\Omega\,d\Omega'
\\[6pt]
&\quad - \frac{1}{4}\rho^2 \iiiint
\big[({\bf u}\cdot\nabla)
({\bf \Omega}\cdot{\bf n}({\bf r}))\big]
f'({\bf \Omega}\cdot{\bf n}({\bf r}))
\\
&\quad \times
U_{MS}({\bf u};{\bf \Omega},{\bf \Omega}')
f'({\bf \Omega}'\cdot{\bf n}({\bf r}))
\\
&\quad \times
\big[({\bf u}\cdot\nabla)
({\bf \Omega}'\cdot{\bf n}({\bf r}))\big]
\,d{\bf r}\,d{\bf u}\,d\Omega\,d\Omega'.
\end{aligned}
\label{A4}
\end{equation}
%\end{widetext}
The first term vanishes because its integrand is odd in $\mathbf{u}$ while $U_{\rm MS}$ is even, as in Straley's theory \cite{Straley73}. The second-order term therefore determines the excess free energy used to derive the Frank elastic constants.

%%%%% Fig.1 %%%%%%%%%%%
\begin{figure}
\begin{center}
\includegraphics*[height=4.5cm]{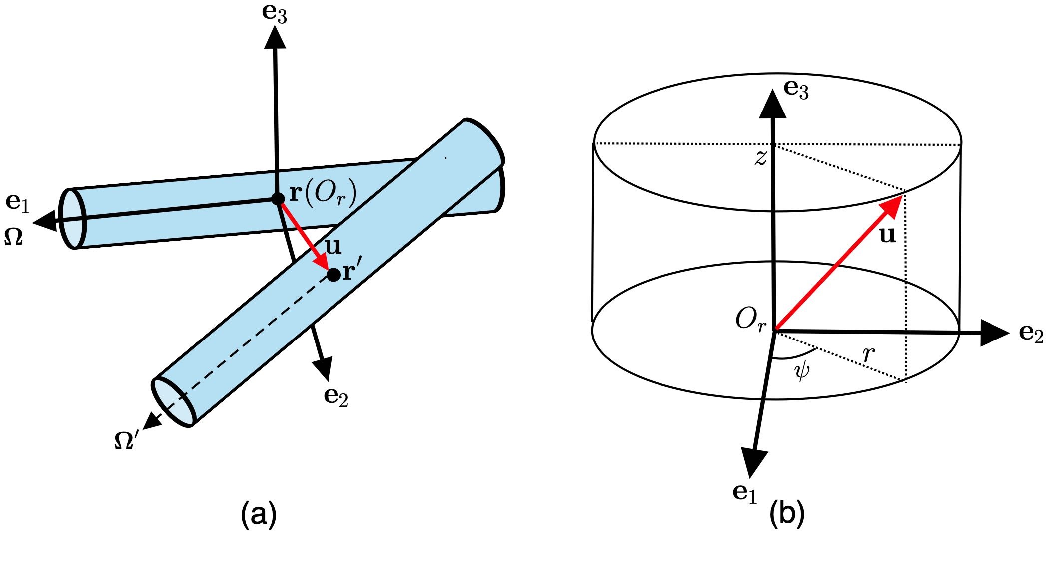}
\caption{Schematic illustration of the anisotropic MS attractive interaction between two rod-like molecules.
(a) The position vectors of two interacting points on the rods are denoted by
${\bf r}$ and ${\bf r}'$, respectively, and the separation vector is
${\bf u}={\bf r}'-{\bf r}$.
The molecular orientations are represented by the unit vectors
${\bf\Omega}$ and ${\bf\Omega}'$, while
${\bf e}_3=
({\bf\Omega}\times{\bf\Omega}')/
|{\bf\Omega}\times{\bf\Omega}'|
$
is the unit vector normal to the intermolecular plane defined by
${\bf\Omega}$ and ${\bf\Omega}'$. (b) The cylindrical-coordinate representation of their separation vector $\mathbf{u}$ in the local orthonormal basis $(\mathbf e_1,\mathbf e_2,\mathbf e_3)$, where we set ${\bf\Omega}$ to be parallel to $\mathbf e_1$. The MS attractive interaction between molecules acts within the cylindrical interaction volume.
}
\label{fig1}
\end{center}
\end{figure}
%%%%%%%%%%%%%%%

The attractive interaction between two rod-like molecules is modeled by the following anisotropic MS potential:
 \begin{eqnarray}
U_{MS}({\bf u}; {\bf\Omega}, {\bf\Omega}')=-\nu_MP_2({\bf\Omega}\cdot {\bf\Omega}') W({\bf u}),
\label{A6}
\end{eqnarray}
where $P_2(\eta)=(3/2)\eta^2-1/2$ is the second Legendre polynomial,
$\nu_M=U_a/(k_BT)$ is the dimensionless interaction strength,
$U_a$ is the characteristic attractive interaction energy,
and $W({\bf u})$ is a window function that confines the attractive interaction to a finite interaction volume.
The anisotropy of the attractive interaction is introduced through the shape of this interaction volume, as described below.

%%%%%%%%%%%%%%%%
%%%%%  %%%%%%%%%%
%%%%%%%%%%%%%%%%
Figure~\ref{fig1} schematically illustrates the anisotropic MS interaction between two rod-like molecules. 
Fig.~\ref{fig1}(a) shows the position vectors of two interacting points on the rods, denoted by ${\bf r}$ and ${\bf r}'$, respectively, and the separation vector is defined as ${\bf u}={\bf r}'-{\bf r}$. 
The molecular orientations are represented by the unit vectors ${\bf\Omega}$ and ${\bf\Omega}'$, while
${\bf e}_3=({\bf\Omega}\times{\bf\Omega}')/|{\bf\Omega}\times{\bf\Omega}'|$
is the unit vector normal to the intermolecular plane defined by ${\bf\Omega}$ and ${\bf\Omega}'$. 
As shown in Fig.~\ref{fig1}(b), the separation vector can be expressed in the local orthonormal basis $(\mathbf e_1,\mathbf e_2,\mathbf e_3)$ associated with cylindrical coordinates as
\begin{eqnarray}
{\bf u}&=& x{\bf e}_1+y{\bf e}_2+z{\bf e}_3
\nonumber \\
&=&r\cos\psi\,{\bf e}_1+r\sin\psi\,{\bf e}_2+z{\bf e}_3,
\label{A8}
\end{eqnarray}
where \(x=r\cos\psi\), \(y=r\sin\psi\), and \(r=\sqrt{x^2+y^2}\).
 We define the orthonormal basis vectors as
\begin{equation}
{\bf e}_1= {\bf \Omega},
\label{A8e1}
\end{equation}
\begin{equation}
{\bf e}_2= {\bf e}_3 \times {\bf e}_1=\frac{{\bf \Omega}'-\cos\gamma_{12} {\bf \Omega}}{|{\bf \Omega}\times {\bf \Omega}'|},
\label{A8e2}
\end{equation}
and
\begin{equation}
{\bf e}_3=\frac{ {\bf \Omega}\times {\bf \Omega}'}{|{\bf \Omega}\times {\bf \Omega}'|},
\label{A8e3}
\end{equation}
with $|{\bf \Omega}\times {\bf \Omega}'| =|\sin\gamma_{12}|$.
The interaction depends on both the relative positions and the orientations of the two molecules, allowing its range to differ within the intermolecular plane and along the normal direction. Here, we assume that the MS attractive interaction acts within a finite cylindrical region, defined by the window function
\begin{equation}
W(\mathbf{u})=
\begin{cases}
1, & 0<r<L_c/2,\quad 0<\psi<2\pi,\quad |z|<L_z/2,\\[2mm]
0, & \text{otherwise},
\end{cases}
\label{A7}
\end{equation}
where $L_c$ denotes the interaction range within the intermolecular plane, whereas $L_z$ specifies the interaction range along the normal direction $\mathbf{e}_3$.

 Using Eq.~(\ref{A8}) together with $d{\bf u}=r dr d\psi   dz$ for the cylindrical coordinates, the integration can be carried out analytically as
%\begin{widetext}
\begin{equation}
\begin{aligned}
&\int U_{MS}({\bf u};{\bf \Omega},{\bf \Omega}')
({\bf u}\cdot{\bf a})({\bf u}\cdot{\bf b})\,d{\bf u}
\\
&= -\frac{1}{16}\nu_M P_2({\bf \Omega}\cdot{\bf \Omega}')
V_{int}
\bigg[
L_c^2({\bf e}_1\cdot{\bf a})({\bf e}_1\cdot{\bf b})
\\
&\qquad
+L_c^2({\bf e}_2\cdot{\bf a})({\bf e}_2\cdot{\bf b})
\\
&\qquad
+\frac{4}{3}L_z^2
({\bf e}_3\cdot{\bf a})({\bf e}_3\cdot{\bf b})
\bigg],
\end{aligned}
\label{A9}
\end{equation}
%\end{widetext}
 for arbitrary vectors ${\bf a}$ and ${\bf b}$, where $V_{int}=(\pi/4)L_c^2L_z$ is the cylindrical interaction volume.
Substituting Eq.~(\ref{A9}) into Eq.~(\ref{A4}), we obtain
%\begin{widetext}
\begin{equation}
\begin{aligned}
\beta \Delta F
&= \frac{L^2}{16\pi D^3}\nu_M \phi^2
\iiint
f'({\bf \Omega}\cdot{\bf n}({\bf r}))
f'({\bf \Omega}'\cdot{\bf n}({\bf r}))
\\
&\quad \times
P_2({\bf \Omega}\cdot{\bf \Omega}')
G_i({\bf e}_1,{\bf e}_2,{\bf e}_3)
\,d{\bf r}\,d\Omega\,d\Omega'.
\end{aligned}
\label{A10}
\end{equation}
%\end{widetext}
where the geometrical factor is given by
\begin{widetext}
\begin{equation}
\begin{aligned}
G_i({\bf e}_1,{\bf e}_2,{\bf e}_3)
&= \alpha^4\left(\frac{L_z}{D}\right)
\bigg[
({\bf e}_1\cdot\nabla)({\bf \Omega}\cdot{\bf n})
({\bf e}_1\cdot\nabla)({\bf \Omega}'\cdot{\bf n})
\\
&\quad+
({\bf e}_2\cdot\nabla)({\bf \Omega}\cdot{\bf n})
({\bf e}_2\cdot\nabla)({\bf \Omega}'\cdot{\bf n})
\\
&\quad+
\frac{4}{3}\left(\frac{L_z}{L_c}\right)^2
({\bf e}_3\cdot\nabla)({\bf \Omega}\cdot{\bf n})
({\bf e}_3\cdot\nabla)({\bf \Omega}'\cdot{\bf n})
\bigg].
\end{aligned}
\label{A11}
\end{equation}
\end{widetext}
The normalized interaction range within the intermolecular plane is defined by
\begin{equation}
\alpha=L_c/L.
\label{LLL}
\end{equation}
The geometrical factors $G_i$ correspond to the splay ($i=1$), twist ($i=2$), and bend ($i=3$) deformations, respectively. Equation~(\ref{A11}) depends only on the spatial gradients of the director field. For example,
$
({\bf e}_1\!\cdot\!\nabla)
({\bf\Omega}\!\cdot\!{\bf n})
=
e_{1\alpha}\Omega_\beta
(\partial_\alpha n_\beta).
$

Equation~(\ref{A10}) is already second order in the director distortions and can therefore be directly compared with the Frank elastic free energy,
\begin{equation}
\begin{aligned}
\Delta F = \frac{1}{2}\int \big[
&K_{11}(\nabla\cdot{\bf n})^2
+K_{22}\big({\bf n}\cdot(\nabla\times{\bf n})\big)^2
\\
&+K_{33}\big({\bf n}\times(\nabla\times{\bf n})\big)^2
\big]\,d{\bf r},
\end{aligned}
\label{A12}
\end{equation}
where $K_{11}$, $K_{22}$, and $K_{33}$ denote the splay, twist, and bend elastic constants, respectively.
Comparison of Eq.~(\ref{A10}) with Eq.~(\ref{A12}) yields
\begin{equation}
\begin{aligned}
K_{ii}
&= \frac{c_0}{8}\nu_M\phi^2
\iint
f'({\bf\Omega}\!\cdot\!{\bf n})
f'({\bf\Omega}'\!\cdot\!{\bf n})
\\
&\quad \times
P_2({\bf\Omega}\!\cdot\!{\bf\Omega}')
G_i({\bf e}_1,{\bf e}_2,{\bf e}_3)
\,d\Omega\,d\Omega',
\end{aligned}
\label{A13}
\end{equation}
where
$c_0=k_BTn^2/(\pi D)$ is a characteristic elastic scale,
$\phi=\pi(D/2)^2L\rho$
is the volume fraction of the rod-like molecules,
and the reduced elastic constants are defined by
$\tilde{K}_{ii}=K_{ii}/c_0$.

We choose the nematic director ${\bf n}$ to be parallel to the $x$-axis, and the molecular orientation is represented by
\begin{eqnarray}
{\bf \Omega}= (\cos\theta, \sin\theta\cos\varphi, \sin\theta\sin\varphi),
\label{A14}
\end{eqnarray}
where $\theta$ and $\varphi$ are the polar and azimuthal angles, respectively. Retaining terms linear in the distortion coordinates \cite{Straley73,MatsuCRC}, we obtain the geometrical factors in Eq.~(\ref{A11}) for the three director deformations.
For the splay ($i=1$) deformation, we take ${\bf n}=(1,y,0)$, yielding
\begin{equation}
\begin{aligned}
G_1^{MS}
&= \alpha^4\left(\frac{L_z}{D}\right)
({\bf \Omega}\cdot{\bf e}_y)
({\bf \Omega}'\cdot{\bf e}_y)
\\
&\quad \times
\bigg[
({\bf e}_1\cdot{\bf e}_y)^2
+({\bf e}_2\cdot{\bf e}_y)^2
+\frac{4}{3}\left(\frac{L_z}{L_c}\right)^2
({\bf e}_3\cdot{\bf e}_y)^2
\bigg]
\\
&= \alpha^4\left(\frac{L_z}{D}\right)
({\bf \Omega}\cdot{\bf e}_y)
({\bf \Omega}'\cdot{\bf e}_y)
\\
&\quad \times
\bigg[
1+A_{cyc}({\bf e}_3\cdot{\bf e}_y)^2
\bigg] ,
\end{aligned}
\label{A_G1}
\end{equation}
where $G_i^{MS}=G_i({\bf e}_1,{\bf  e}_2,{\bf e}_3)$ and we define the shape-anisotropy parameter 
\begin{equation}
A_{cyc}=\frac{4}{3} \big(\frac{L_z}{L_c}\big)^2-1,
\label{A_cyc_1}
\end{equation}
which controls the anisotropic part of the deformation-dependent geometrical factors. Here, we use the relation $e_{1\alpha}^2+e_{2\alpha}^2+e_{3\alpha}^2=1$ ($\alpha=x, y, z$).
For the twist ($i=2$) deformation, we take ${\bf n}=(1,z,0)$, yielding
\begin{equation}
\begin{aligned}
G_2^{MS}
&= \alpha^4\left(\frac{L_z}{D}\right)
({\bf \Omega}\cdot{\bf e}_y)
({\bf \Omega}'\cdot{\bf e}_y)
\\
&\quad\times
\bigg[
({\bf e}_1\cdot{\bf e}_z)^2
+({\bf e}_2\cdot{\bf e}_z)^2
+\frac{4}{3}\left(\frac{L_z}{L_c}\right)^2
({\bf e}_3\cdot{\bf e}_z)^2
\bigg]
\\
&= \alpha^4\left(\frac{L_z}{D}\right)
({\bf \Omega}\cdot{\bf e}_y)
({\bf \Omega}'\cdot{\bf e}_y)
\\
&\quad\times
\bigg[
1+A_{cyc}({\bf e}_3\cdot{\bf e}_z)^2
\bigg].
\end{aligned}
\label{A_G2}
\end{equation}
For the bend ($i=3$) deformation, we take ${\bf n}=(1,x,0)$, yielding
\begin{equation}
\begin{aligned}
G_3^{MS}
&= \alpha^4\left(\frac{L_z}{D}\right)
({\bf \Omega}\cdot{\bf e}_y)
({\bf \Omega}'\cdot{\bf e}_y)
\\
&\quad\times
\bigg[
({\bf e}_1\cdot{\bf e}_x)^2
+({\bf e}_2\cdot{\bf e}_x)^2
+\frac{4}{3}\left(\frac{L_z}{L_c}\right)^2
({\bf e}_3\cdot{\bf e}_x)^2
\bigg]
\\
&= \alpha^4\left(\frac{L_z}{D}\right)
({\bf \Omega}\cdot{\bf e}_y)
({\bf \Omega}'\cdot{\bf e}_y)
\\
&\quad\times
\bigg[
1+A_{cyc}({\bf e}_3\cdot{\bf e}_x)^2
\bigg].
\end{aligned}
\label{A_G3}
\end{equation}
Equations~(\ref{A_G1})--(\ref{A_G3}) show that the elastic anisotropy originates from the competition between attractive interactions within the intermolecular plane and those along its normal direction ${\bf e}_3$. Their relative contributions are controlled by the shape-anisotropy parameter $A_{\rm cyc}$.

%%%%%%%%%%%%%%%%%%%%%%%%%%%%%%%%%%%
\section{Numerical Results and Discussion}
\label{sec3}
%%%%%%%%%%%%%%%%%%%%%%%%%%%%%%%%%%%

Using the microscopic expressions for the Frank elastic constants (Eq.~(\ref{A13})) derived in the previous section, we numerically evaluate the elastic constants as functions of concentration and temperature. For the numerical calculations, we employ the orientational distribution function of rod-like molecules obtained from the Maier--Saupe molecular-field theory \cite{Matsu96,MatsuCRC}:
\begin{equation}
f(\eta=\cos\theta)=\frac{1}{Z_0}
\exp\!\left(
\nu_M n\phi S P_2(\eta)
\right),
\label{b1}
\end{equation}
where $Z_0$ is the normalization constant satisfying
$
\int f(\eta)\,d\Omega=1.
$
The derivative of the distribution function is given by
$
f'(\eta)=3\nu_M n\phi S\,\eta\,f(\eta).
$
The orientational order parameter $S$ is determined self-consistently from
\begin{eqnarray}
S
&=&
\int P_2(\eta)f(\eta)\,d\Omega
\nonumber\\
&=&
\frac{I_1(S)}{I_0(S)},
\label{b3}
\end{eqnarray}
where
\begin{equation}
I_q(S)
=
\int_0^1
\left[P_2(\eta)\right]^q
\exp\!\left(
\nu_M n\phi S P_2(\eta)
\right)
d\eta,
\label{b4}
\end{equation}
with $q=0$ and $1$. The normalization constant is therefore given by
$
Z_0=4\pi I_0(S).
$

%%%%%%%%%%%%%%%%%%%%%%%%%%%%%%%%%%%
\subsection{Concentration Dependence of the Frank Elastic Constants}
\label{conc}
%%%%%%%%%%%%%%%%%%%%%%%%%%%%%%%%%%%

%%%%% Fig.2 %%%%%%%%%%%
\begin{figure*}[t]
\begin{center}
\includegraphics*[height=5.cm]{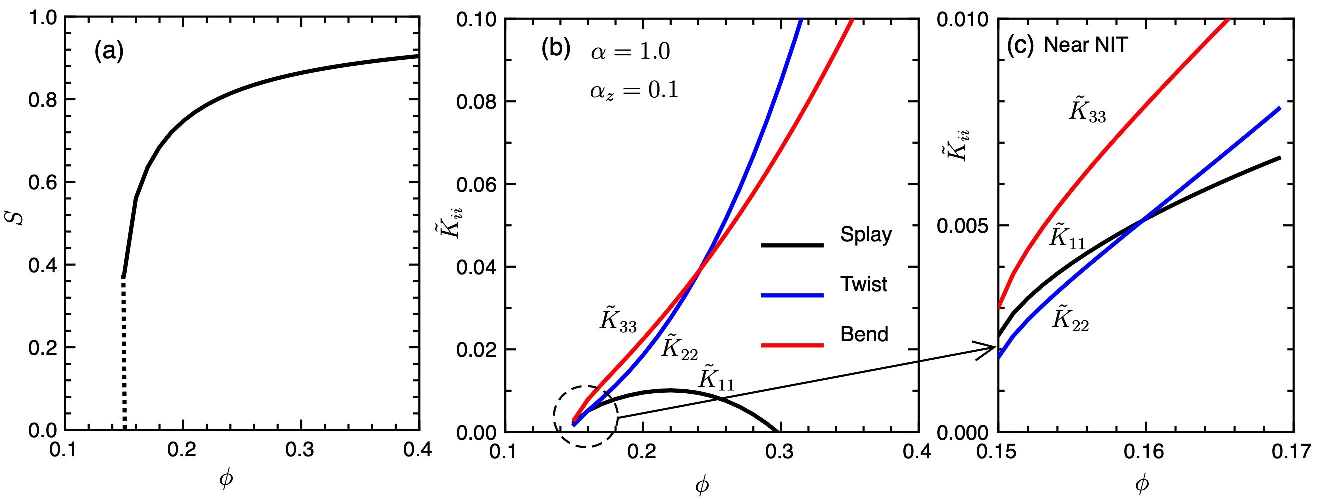}
\caption{
(a) Concentration dependence of the nematic order parameter $S$.
(b) Reduced Frank elastic constants $\tilde{K}_{11}$,
$\tilde{K}_{22}$, and $\tilde{K}_{33}$ as functions of the
volume fraction $\phi$ for
$\alpha_z=0.1$.
(c) Reduced Frank elastic constants $\tilde{K}_{11}$,
$\tilde{K}_{22}$, and $\tilde{K}_{33}$ near the NIT concentration.
}
\label{fig2}
\end{center}
\end{figure*}
%%%%%%%%%%%%%%%

Here, we show the concentration dependence of the Frank elastic constants.
Throughout this section, the interaction range within the intermolecular plane is fixed at
$\alpha=L_c/L=1$, while the interaction range along the normal to the intermolecular plane is varied through the dimensionless parameter
\begin{equation}
\alpha_z=\frac{L_z}{L_c}=\frac{L_z}{L},
\label{alpha_z}
\end{equation}
where
$L_z/D=\alpha_zn$
in Eqs.~(\ref{A_G1})--(\ref{A_G3}).
Thus, increasing $\alpha_z$ corresponds to extending the attractive interaction along the normal direction while keeping the in-plane interaction range unchanged.
We set
$\nu_M=3.0$
and an aspect ratio
$n=L/D=10$.

Figure~\ref{fig2} shows the concentration $\phi$ dependence of the nematic order parameter (a) and the reduced Frank elastic constants (b), (c) for $\alpha_z=0.1$, corresponding to a flattened cylindrical interaction volume that is thin along the $\mathbf{e}_3$ direction. As shown in Fig.~2(a), the nematic order parameter $S$ exhibits a discontinuous jump at the nematic--isotropic transition (NIT) at $\phi \simeq 0.15$ and increases with increasing volume fraction $\phi$ of the rods. The corresponding reduced elastic constants $\widetilde{K}_{ii}$ are shown in Figs.~2(b) and (c).
Near the NIT, the bend elastic constant is the largest, followed by the splay and twist elastic constants, i.e., $K_{33}>K_{11}>K_{22}$, as shown in Fig.~\ref{fig2}(c). For a small $\alpha_z$, the anisotropic MS interaction yields the same hierarchy of the Frank elastic constants near the NIT as that found in Straley's excluded-volume model \cite{Straley73}. The elastic constants $K_{11}$ and $K_{22}$ cross at $\phi_1 \simeq 0.16$. We also find that $K_{33}$ and $K_{22}$ cross at $\phi_2 \simeq 0.24$, as shown in Fig.~\ref{fig2}(b). At high concentrations, $\widetilde{K}_{11}$ displays a non-monotonic concentration dependence, and the hierarchy becomes $K_{22}>K_{33}>K_{11}$.
For the strongly oblate interaction volume with $\alpha_z=0.1$, the interaction range along the $\mathbf{e}_3$ direction is much shorter than that in the transverse directions. Consequently, the contribution associated with spatial variations along $\mathbf{e}_3$ is strongly suppressed. This pronounced anisotropy of the interaction volume leads to distinct contributions to the splay and twist geometrical factors, as described in Appendix~\ref{app:strong-alignment}.

%%%%% Fig.3 %%%%%%%%%%%
\begin{figure*}[t]
\begin{center}
\includegraphics*[height=11cm]{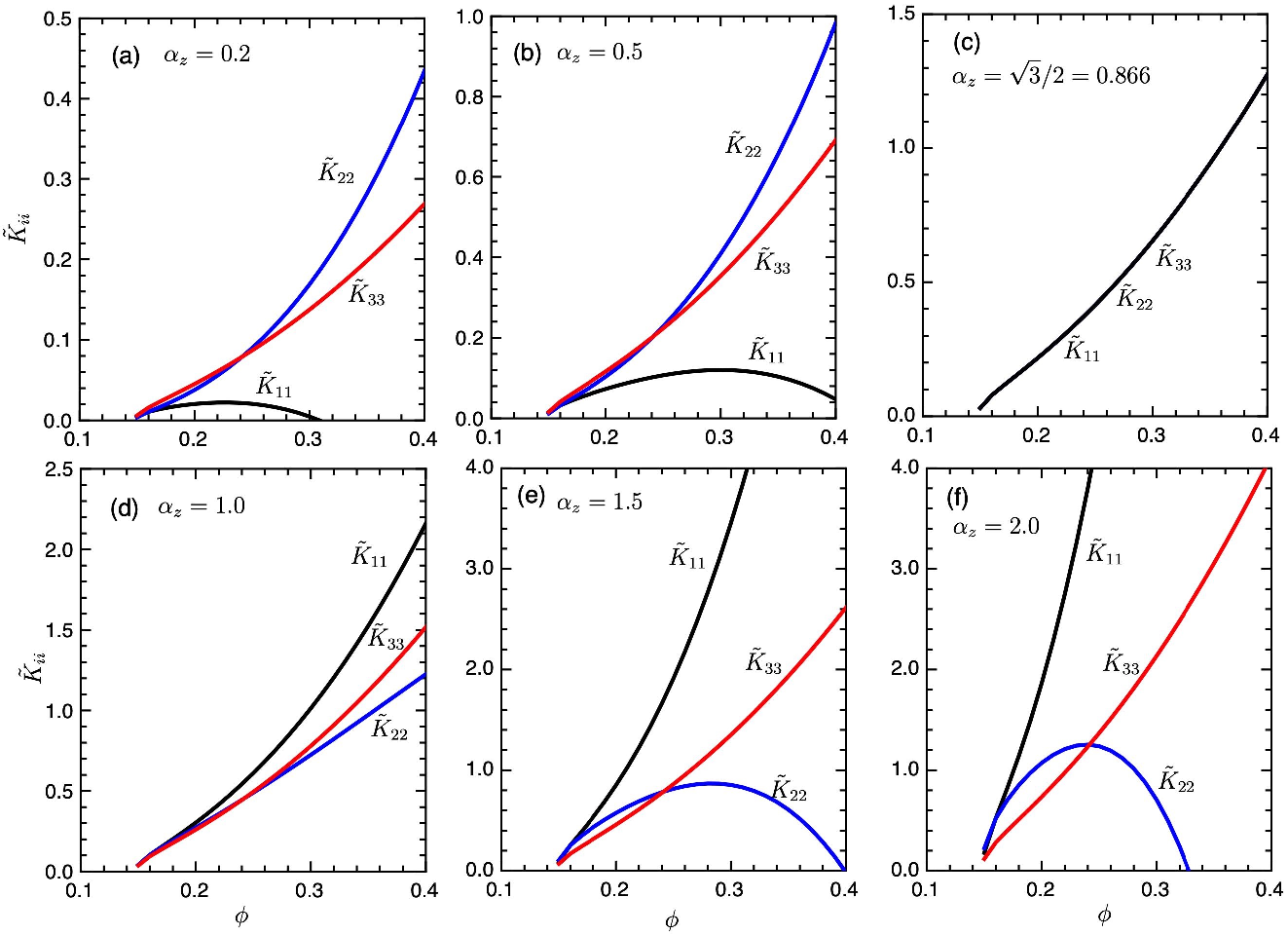}
\caption{Reduced Frank elastic constants $\widetilde{K}_{11}$ (splay, black), $\widetilde{K}_{22}$ (twist, blue), and $\widetilde{K}_{33}$ (bend, red) as functions of the volume fraction $\phi$ for cylindrical interaction volumes with different aspect ratios $\alpha_z=L_z/L_c$: (a) $\alpha_z=0.2$, (b) $\alpha_z=0.5$, (c) $\alpha_z=\sqrt{3}/2\simeq0.866$, (d) $\alpha_z=1.0$, (e) $\alpha_z=1.5$, and (f) $\alpha_z=2.0$. At $\alpha_z=\sqrt{3}/2$, the three elastic constants are identical because the transverse and axial second moments of the interaction volume are equal.
}
\label{fig3}
\end{center}
\end{figure*}
%%%%%%%%%%%%%%%
Figure~\ref{fig3} shows the concentration dependence of the three reduced Frank elastic constants for various values of the interaction-volume aspect ratios $\alpha_z$.
For a cylindrical interaction volume with diameter $L_c$ and height $L_z$, the transverse and axial second moments are given by
\begin{equation}
M_\perp=\int W(\mathbf{u})x^2\,d\mathbf{u}
=\int W(\mathbf{u})y^2\,d\mathbf{u}
=V_{\rm int}\frac{L_c^2}{16},
\label{M_}
\end{equation}
and
\begin{equation}
M_z=\int W(\mathbf{u})z^2\,d\mathbf{u}
=V_{\rm int}\frac{L_z^2}{12}.
\label{M_1}
\end{equation}
Their ratio is given by
\begin{equation}
\frac{M_z}{M_\perp}
=\frac{4}{3}\left(\frac{L_z}{L_c}\right)^2
=\frac{4}{3}\alpha_z^2=A_{cyc}+1,
\label{M_z11}
\end{equation}
where \(A_{\rm cyc}\) is the shape-anisotropy parameter defined in Eq.~(\ref{A_cyc_1}).
The anisotropic terms in Eqs.~(\ref{A_G1})--(\ref{A_G3}) therefore modify the geometrical factors through $\alpha_z$.

For $\alpha_z<\sqrt{3}/2$, the interaction volume is oblate in terms of its second moments ($M_z<M_\perp$) and $A_{\rm cyc}<0$ [Figs.~3(a) and 3(b)]. 
The behavior near the NIT is shown in Fig.~\ref{fig2}(c). As the volume fraction increases, the differences among the three elastic modes become more pronounced. In particular, $K_{11}$ decreases and eventually becomes negative, indicating a splay instability of the uniform nematic state within the present approximation. At high concentrations, the elastic constants satisfy
$
K_{22}>K_{33}>K_{11}.
$

At the special aspect ratio $\alpha_z=\sqrt{3}/2\simeq 0.866$ [Fig.~3(c)], the geometrical coefficient is $A_{\rm cyc}=0$ in Eqs.~(\ref{A_G1})--(\ref{A_G3}), and the transverse and axial second moments of the cylindrical interaction volume are identical:
\begin{equation}
M_\perp=V_{\rm int}\frac{L_c^2}{16}
=V_{\rm int}\frac{L_z^2}{12}
=M_z.
\label{MMM}
\end{equation}
Although the interaction volume remains geometrically cylindrical, it is isotropic at the level of the second spatial moments entering the elastic free energy. We then obtain
$
K_{11}=K_{22}=K_{33},
$
where the one-constant approximation is valid at $\alpha_z=\sqrt{3}/2$ within the present model. This equality holds throughout the nematic concentration range.

For $\alpha_z>\sqrt{3}/2$, or equivalently $A_{\rm cyc}>0$, the cylindrical interaction volume is prolate in terms of its second moments, such that $M_z>M_\perp$. This geometrical anisotropy is reflected in the relative magnitudes of the Frank elastic constants at high concentrations, where we obtain
$
K_{11}>K_{33}>K_{22}.
$
For $\alpha_z=1.0$ [Fig.~3(d)], all three elastic constants increase with increasing volume fraction, with $\widetilde{K}_{11}$ exhibiting the most rapid growth. As the interaction volume becomes more elongated, i.e., for $\alpha_z=1.5$ and $2.0$ [Figs.~3(e) and 3(f)], the splay elastic constant is further enhanced and becomes much larger than the bend and twist elastic constants. In contrast, $\widetilde{K}_{22}$ exhibits a pronounced non-monotonic concentration dependence: it initially increases, reaches a maximum, and subsequently decreases, becoming negative at high volume fractions. 
Negative values of a Frank elastic constant indicate that the uniform nematic state is unstable to the corresponding director deformation. Determining the resulting state would require higher-order gradient terms, which are not included in the present analysis.

%%%%% %%%%%%%%%%%%%%%% %%%%%%%%%%%%%%%% %%%%%%%%%%%
%%%%% Fig.4 %%%%%%%%%%%
\begin{figure*}[t]
\begin{center}
\includegraphics*[height=6cm]{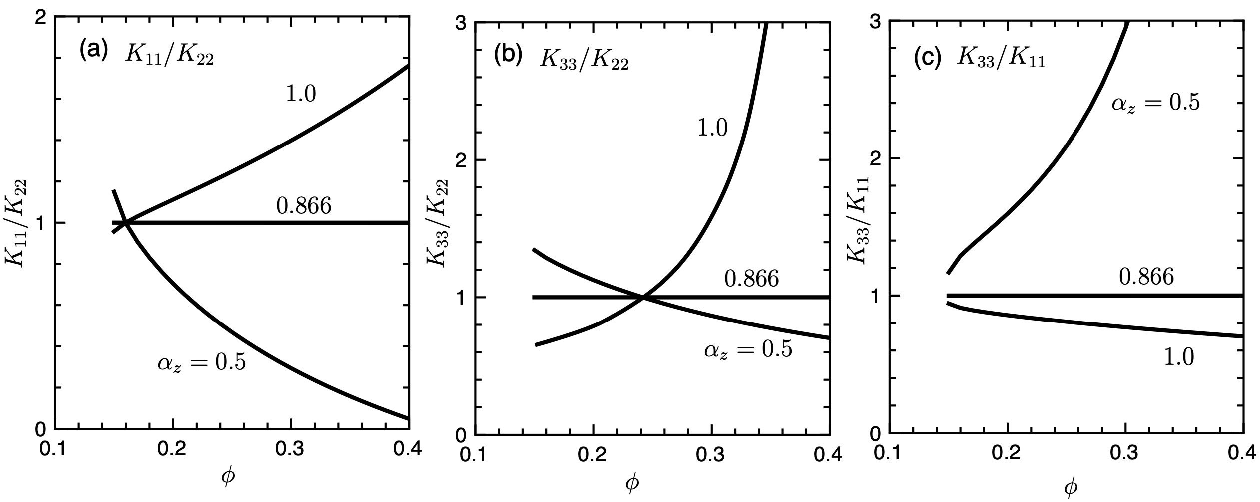}
\caption{Concentration dependence of the ratios of the Frank elastic constants for cylindrical interaction volumes with $\alpha_z=0.5$, $\alpha_z=\sqrt{3}/2\simeq 0.866$, and $\alpha_z=1.0$: (a) $K_{11}/K_{22}$, (b) $K_{33}/K_{22}$, and (c) $K_{33}/K_{11}$. At $\alpha_z=\sqrt{3}/2$, all three ratios are unity throughout the nematic concentration range.
}
\label{fig4}
\end{center}
\end{figure*}
%%%%%%%%%%%%%%%

The decrease of $K_{22}$ for large $\alpha_z$ can be understood from Eq.~(\ref{A_G2}). In the strongly aligned limit, with $\mathbf{n}=\mathbf{e}_x$, we have $(\mathbf{e}_3\cdot\mathbf{e}_z)^2\simeq(\Omega'_y-\Omega_y)^2/(\Delta\Omega)^2$, as shown in Appendix A, where $(\Delta\Omega)^2=(\Omega'_y-\Omega_y)^2+(\Omega'_z-\Omega_z)^2$. Thus, the anisotropic contribution proportional to $A_{\rm cyc}\Omega_y\Omega'_y(\mathbf{e}_3\cdot\mathbf{e}_z)^2$ gives greater weight to molecular pairs with a large difference in their $y$ components. For a strongly aligned and azimuthally symmetric orientational distribution, such pairs can have oppositely signed transverse components, giving $\Omega_y\Omega'_y<0$. Consequently, for $\alpha_z>\sqrt{3}/2$, where $A_{\rm cyc}>0$, this anisotropic contribution tends to reduce $K_{22}$. As the volume fraction increases, this contribution becomes increasingly important, resulting in the non-monotonic concentration dependence of $\widetilde{K}_{22}$.

Figure~\ref{fig4} shows the concentration dependence of the ratios between the Frank elastic constants for oblate ($\alpha_z=0.5$), isotropic-second-moment ($\alpha_z=\sqrt{3}/2$), and prolate ($\alpha_z=1.0$) cylindrical interaction volumes.
In Fig.~\ref{fig4}(a), the ratio $K_{11}/K_{22}$ decreases with increasing $\phi$ for $\alpha_z=0.5$, indicating that the splay elastic constant decreases relative to the twist elastic constant. In contrast, for $\alpha_z=1.0$, $K_{11}/K_{22}$ increases monotonically, indicating that the splay elastic constant increases relative to the twist elastic constant. 
For $\alpha_z=0.5$ and $1.0$, $K_{11}/K_{22}$ crosses unity at the same volume fraction, $\phi_1\simeq0.16$; at $\alpha_z=\sqrt{3}/2$, the ratio remains unity for all $\phi$.

Figure~\ref{fig4}(b) shows the ratio $K_{33}/K_{22}$ as a function of $\phi$. For $\alpha_z=0.5$, this ratio decreases with increasing concentration and falls below unity, whereas for $\alpha_z=1.0$, it increases. Thus, for an oblate interaction volume ($\alpha_z=0.5$), the bend elastic constant becomes smaller relative to the twist elastic constant, whereas for a prolate interaction volume ($\alpha_z=1.0$), it becomes substantially larger. At $\alpha_z=\sqrt{3}/2$, the two elastic constants are equal. We also find that $K_{33}/K_{22}=1$ at $\phi_2\simeq0.24$.

Figure~\ref{fig4}(c) shows the ratio $K_{33}/K_{11}$. For $\alpha_z=0.5$, the ratio increases with $\phi$, whereas for $\alpha_z=1.0$, it remains below unity. Thus, for the oblate interaction volume, the bend elastic constant increases relative to the splay elastic constant with increasing concentration, while for the prolate interaction volume, the splay elastic constant remains larger than the bend elastic constant.

%%%%%%%%%%%%%%%%%%%%%%%%%%%%%%%%%
\subsection{Temperature dependence of the Frank Elastic Constants}
\label{temp}
%%%%%%%%%%%%%%%%%%%%%%%%%%%%%%%%%

The concentration dependence discussed above is relevant to lyotropic nematic liquid crystals, whereas the temperature dependence at a fixed volume fraction $\phi=1$ is relevant to thermotropic nematic liquid crystals. In the following, we introduce the reduced temperature
\[
\tau=\frac{k_BT}{U_a}=\frac{1}{\nu_M},
\]
and set the aspect ratio of the rods to $n=L/D=3$.

%%%%% Fig.5 %%%%%%%%%%%
\begin{figure*}[t]
\begin{center}
\includegraphics*[height=6cm]{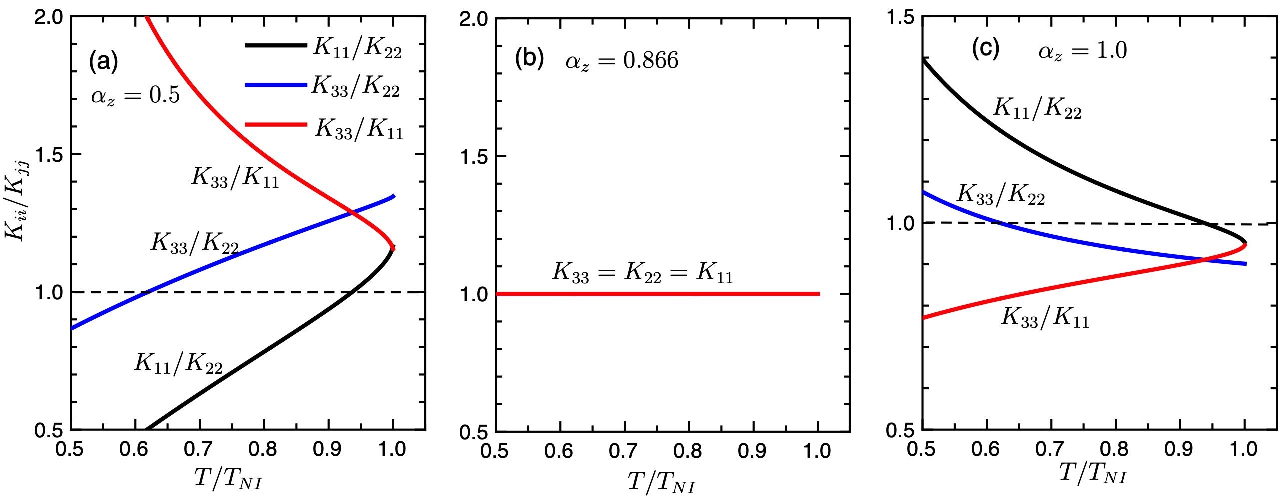}
\caption{Temperature dependence of the ratios of the Frank elastic constants for cylindrical interaction volumes with (a) $\alpha_z=0.5$, (b) $\alpha_z=\sqrt{3}/2\simeq0.866$, and (c) $\alpha_z=1.0$. The black, blue, and red curves represent $K_{11}/K_{22}$, $K_{33}/K_{22}$, and $K_{33}/K_{11}$, respectively. The horizontal dashed lines indicate unity. At $\alpha_z=\sqrt{3}/2$, all three ratios are identically equal to unity, corresponding to the one-constant relation $K_{11}=K_{22}=K_{33}$.
}
\label{fig5}
\end{center}
\end{figure*}
%%%%%%%%%%%%%%%
Figure~\ref{fig5} shows the ratios $K_{ii}/K_{jj}$ between the Frank elastic constants as functions of the reduced temperature $T/T_{\rm NI}$, where $T_{\rm NI}$ is the nematic--isotropic transition (NIT) temperature, for cylindrical interaction volumes with different aspect ratios $\alpha_z$.

For the oblate interaction volume with $\alpha_z=0.5$ [Fig.~\ref{fig5}(a)], the ratio $K_{33}/K_{11}$ remains larger than unity over the entire temperature range shown. Thus, the relative ordering of $K_{33}$ and $K_{11}$ does not change, whereas $K_{11}/K_{22}$ and $K_{33}/K_{22}$ cross unity at $T_1/T_{\rm NI}\simeq0.94$ and $T_2/T_{\rm NI}\simeq0.62$, respectively.
Near $T_{\rm NI}$, the twist elastic constant $K_{22}$ is the smallest, and the hierarchy is $K_{33}>K_{11}>K_{22}$. As the temperature decreases, $K_{22}$ becomes larger than the other two elastic constants. At low temperatures, $T<T_2$, the hierarchy becomes $K_{22}>K_{33}>K_{11}$. The elastic-constant hierarchy consequently changes as
\begin{widetext}
\begin{equation}
K_{22}>K_{33}>K_{11}
\;\overset{T_2(<T_1)}{\longleftrightarrow}\;
K_{33}>K_{22}>K_{11}
\;\overset{T_1(<T_{\rm NI})}{\longleftrightarrow}\;
K_{33}>K_{11}>K_{22}.
\label{K123123}
\end{equation}
\end{widetext}

At the special aspect ratio $\alpha_z=\sqrt{3}/2\simeq0.866$ [Fig.~\ref{fig5}(b)], all ratios are equal to unity, and we have
\begin{equation}
K_{11}=K_{22}=K_{33}.
\label{one-const_1}
\end{equation}
This confirms the validity of the one-constant approximation independently of temperature.

For the prolate interaction volume with $\alpha_z=1.0$ [Fig.~\ref{fig5}(c)], the ratio $K_{33}/K_{11}$ remains below unity over the entire temperature range shown. The twist elastic constant $K_{22}$ crosses the bend and splay elastic constants at $T_2$ and $T_1$, respectively, leading to the following sequence of elastic-constant orderings:
\begin{widetext}
\begin{equation}
K_{11}>K_{33}>K_{22}
\;\overset{T_2(<T_1)}{\longleftrightarrow}\;
K_{11}>K_{22}>K_{33}
\;\overset{T_1(<T_{NI})}{\longleftrightarrow}\;
K_{22}>K_{11}>K_{33}.
\label{k123_az_big}
\end{equation}
\end{widetext}
Taken together, Eqs.~(\ref{K123123}) and (\ref{k123_az_big}) show that the oblate and prolate interaction volumes exhibit all six possible strict orderings of the three Frank elastic constants. The same sequence of changes in the elastic-constant hierarchy is also found at the characteristic volume fractions $\phi_1(>\phi_{NI})$ and $\phi_2(>\phi_1)$, as discussed in Sec.~\ref{conc}.

In the calculations presented so far, increasing $\alpha_z$ also increases the interaction volume. The absolute values of the Frank elastic constants increase with the interaction volume. To separate the effects of the interaction volume and the interaction anisotropy, we impose the condition of a constant interaction volume:
\begin{equation}
V_{\rm int}=\frac{\pi}{4}L_c^2L_z
=\frac{\pi}{4}L^2D=\mathrm{const.}
\end{equation}
Using $n=L/D$, $\alpha=L_c/L$, and $\alpha_z=L_z/L_c$, this condition becomes
\begin{equation}
n\alpha^3\alpha_z=1.
\end{equation}
Thus, we obtain
\begin{equation}
\alpha=\left(\frac{1}{n\alpha_z}\right)^{1/3}.
\end{equation}
Under this constraint, varying $\alpha_z$ changes only the shape anisotropy of the interaction volume, while keeping its volume constant.

%%%%% Fig.6 %%%%%%%%%%%
\begin{figure*}[t]
\begin{center}
\includegraphics*[height=6cm]{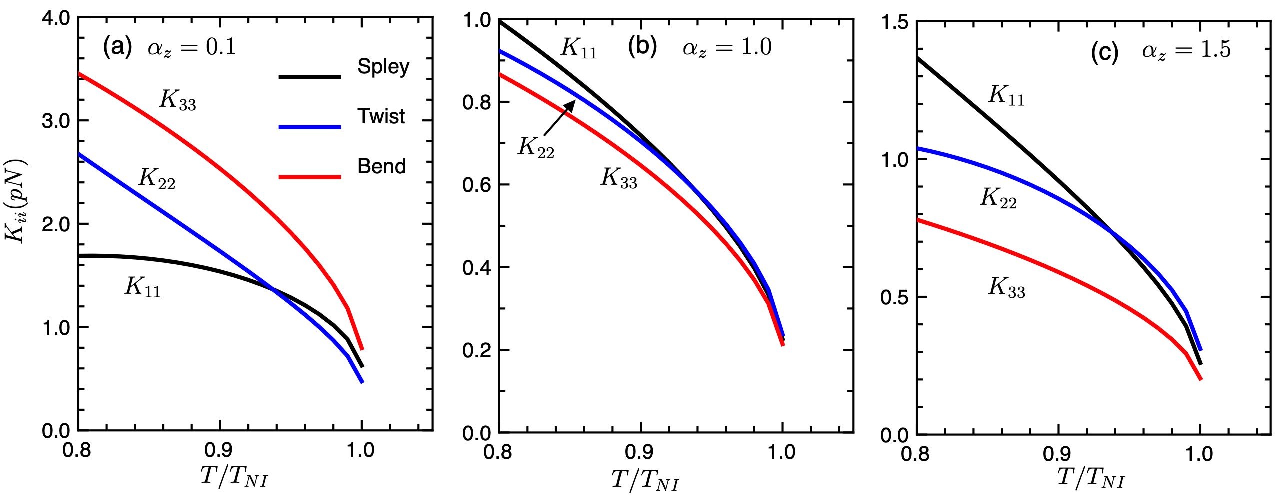}
\caption{Temperature dependence of the Frank elastic constants in physical units near the NIT temperature for cylindrical interaction volumes with (a) $\alpha_z=0.1$, (b) $\alpha_z=1.0$, and (c) $\alpha_z=1.5$, under the constraint of constant interaction volume, $V_{\rm int}=\pi L^2D/4$. The splay ($K_{11}$), twist ($K_{22}$), and bend ($K_{33}$) elastic constants are represented by black, blue, and red curves, respectively. The temperature is normalized by the NIT temperature, $T_{\rm NI}=310~\mathrm{K}$.
}
\label{fig6}
\end{center}
\end{figure*}
%%%%%%%%%%%%%%%
To express the calculated elastic constants in physical units, we set the NIT temperature to $T_{\rm NI}=310~\mathrm{K}$. The molecular diameter is set to $D=2.0~\mathrm{nm}$, and the aspect ratio is fixed at $n=L/D=3$. With these parameters, the characteristic interaction energy is determined to be $U_a=3.86~\mathrm{kJ\,mol^{-1}}$ from the reduced NIT temperature $\tau_{\rm NI}=k_{\rm B}T_{\rm NI}/U_a=0.668$. The conversion factor is then given by
\begin{equation}
c_0(T)=\frac{k_{\rm B}Tn^2}{\pi D}
=6.13\left(\frac{T}{T_{\rm NI}}\right)\mathrm{pN},
\label{c_0c_0}
\end{equation}
and the physical elastic constants are obtained as
$K_{ii}(T)=c_0(T)\widetilde{K}_{ii}(T)$ pN.

Figure~\ref{fig6} shows the temperature dependence of the Frank elastic constants in physical units $\it{near}$ the NIT temperature. 
Although the conversion factor \(c_0(T)\) increases linearly with temperature, the decrease in the reduced elastic constants $\widetilde{K}_{ii}$ is sufficiently large that the physical Frank elastic constants decrease as the temperature approaches \(T_{\rm NI}\).

For the strongly oblate interaction volume with $\alpha_z=0.1$ [Fig.~\ref{fig6}(a)], the bend elastic constant remains larger than the splay elastic constant over the entire temperature range shown. The splay and twist elastic constants intersect at $T_1/T_{\rm NI}\simeq 0.94$. Above $T_1$, the elastic constants are ordered as
$
K_{33}>K_{11}>K_{22}.
$
Below $T_1$, the ordering changes to
$
K_{33}>K_{22}>K_{11}.
$
This behavior is consistent with the oblate case shown in Fig.~\ref{fig5}(a) and the sequence of elastic-constant orderings given in Eq.~(30).

For the prolate interaction volume with $\alpha_z=1.0$ [Fig.~\ref{fig6}(b)], the splay elastic constant is the largest at the lower end of the temperature range shown, followed by the twist and bend elastic constants:
$
K_{11}>K_{22}>K_{33}.
$
As the temperature approaches $T_{\rm NI}$, the twist elastic constant overtakes the splay elastic constant, and the ordering changes to
$
K_{22}>K_{11}>K_{33}.
$
The special value $\alpha_z=\sqrt{3}/2\simeq 0.866$ yields
$K_{11}=K_{22}=K_{33}$, as shown in Fig.~\ref{fig5}(b).
Since $\alpha_z=1.0$ is close to this value, the differences among the three elastic constants remain relatively small.
At lower temperatures, outside the range shown in Fig.~\ref{fig6}(b), the ordering changes to $K_{11}>K_{33}>K_{22}$, as also shown in Fig.~\ref{fig5}(c).

For the more elongated interaction volume with $\alpha_z=1.5$ [Fig.~\ref{fig6}(c)], the splay elastic constant is larger than the bend elastic constant throughout the temperature range shown. At the lower end of this range, the ordering is
$
K_{11}>K_{22}>K_{33}.
$
As the temperature approaches the NIT temperature, the splay and twist elastic constants intersect at $T_1/T_{\rm NI}\simeq 0.94$. Above this temperature, the ordering becomes
$
K_{22}>K_{11}>K_{33}.
$
Thus, the bend elastic constant remains the smallest throughout the temperature range shown. This behavior is consistent with the prolate-case sequence given by Eq.~(\ref{k123_az_big}).
At lower temperatures, outside the range shown in Fig. \ref{fig6}(c), the bend and twist elastic constants intersect at \(T_2\). A similar intersection is shown for \(\alpha_z=1.0\) in Fig.~\ref{fig5}(c).
This intersection completes the sequence of elastic-constant orderings described by Eqs.~(\ref{K123123}) and (\ref{k123_az_big}). The calculated Frank elastic constants are consistent with the typical magnitudes of Frank elastic constants in thermotropic nematic liquid crystals \cite{HAKEMI83,bogi01,Bunning81}.

The hierarchy of the Frank elastic constants changes with temperature and concentration. Near the NIT, the ratio $K_{11}/K_{22}$ crosses unity, thereby reversing the relative magnitudes of the splay and twist elastic constants. At lower temperatures, the ratio $K_{33}/K_{22}$ also crosses unity, thereby reversing the relative magnitudes of the bend and twist elastic constants. In addition, the relative magnitudes of the splay and bend elastic constants depend on the shape of the interaction volume: $K_{33}>K_{11}$ for $\alpha_z<\sqrt{3}/2$, whereas $K_{11}>K_{33}$ for $\alpha_z>\sqrt{3}/2$. These results demonstrate that the hierarchy of the Frank elastic constants is controlled by both the shape of the interaction volume and the degree of nematic ordering.
The geometrical origin of the elastic anisotropy is further elucidated by the strong-alignment approximation derived in Appendix~\ref{app:strong-alignment}.

In this paper, we use a cylindrical interaction volume. The same argument can be extended to a rectangular interaction volume. For a cuboid with a square cross section, $L_x=L_y=L_c$, the second moments are
\begin{equation}
\langle x^2\rangle=\langle y^2\rangle=\frac{L_c^2}{12},
\qquad
\langle z^2\rangle=\frac{L_z^2}{12}.
\label{cuboid_1}
\end{equation}
Therefore, the one-constant condition is obtained at $L_z=L_c$, namely $\alpha_z=L_z/L_c=1$. This difference from the cylindrical value $\alpha_z=\sqrt{3}/2$ indicates that the condition for $K_{11}=K_{22}=K_{33}$ is determined by the second spatial moments of the interaction volume rather than by its shape alone.

%%%%%%%%%%%%%%%%%%%%%%%%%%%%%%%%%%%
\section{Summary}
\label{sec4}
%%%%%%%%%%%%%%%%%%%%%%%%%%%%%%%%%%%
We have theoretically studied the Frank elastic constants of a nematic phase with an anisotropic MS interaction confined to a cylindrical interaction volume. The shape of this volume is characterized by the aspect ratio $\alpha_z=L_z/L_c$. The transverse and axial second moments are equal at the special aspect ratio $\alpha_z=\sqrt{3}/2$, for which the shape-anisotropy parameter $A_{cyc}$ vanishes. At this aspect ratio, the three Frank elastic constants are identical,
$
K_{11}=K_{22}=K_{33},
$
and the one-constant approximation is valid throughout the concentration and temperature ranges considered.

For oblate and prolate interaction volumes, the three elastic constants become distinct. Oblate interaction volumes with $\alpha_z<\sqrt{3}/2$ are characterized by a larger bend elastic constant than the splay elastic constant ($K_{33}>K_{11}$), whereas prolate interaction volumes with $\alpha_z>\sqrt{3}/2$ exhibit the opposite relation ($K_{11}>K_{33}$). Furthermore, the elastic-constant hierarchy is not determined by the interaction-volume shape alone. Increasing the concentration or lowering the temperature strengthens nematic ordering and can change the relative magnitudes of $K_{11}$, $K_{22}$, and $K_{33}$, as shown in Eqs.~(\ref{K123123}) and (\ref{k123_az_big}). The ratios between elastic constants can cross unity, demonstrating concentration- and temperature-induced reversals of the elastic anisotropy.

Using representative molecular parameters, the calculated elastic constants are on the order of several pN, comparable to typical values for thermotropic nematic liquid crystals. These results show that anisotropic attractive interactions play an important role in determining both the magnitude and hierarchy of the Frank elastic constants, providing a simple molecular mechanism for elastic anisotropy in nematic liquid crystals. 
The present model offers a molecular description of anisotropic attractions between rod-like molecules and can be extended to bent-core liquid crystals \cite{matsu19,Ferrarini21}. An important direction for future work is the development of a unified molecular theory that incorporates both excluded-volume interactions and anisotropic MS attractions.

%%%%%%%%%%%
\appendix
\section{Strong-alignment approximation of the geometrical factors}
\label{app:strong-alignment}

In this Appendix, we derive the strong-alignment approximation for the geometrical factors in Eqs.~(\ref{A_G1})--(\ref{A_G3}), which enter the expressions for the Frank elastic constants. We choose the nematic director along the $x$ axis,
\begin{equation}
\mathbf{n}=\mathbf{e}_x.
\label{nnn}
\end{equation}
In the strongly aligned regime, the transverse components of the molecular orientations are small. The orientation vectors can be written as
\begin{equation}
\begin{aligned}
\boldsymbol{\Omega}
&=
\left(
\sqrt{1-\Omega_y^2-\Omega_z^2},
\,\Omega_y,\,\Omega_z
\right)
\\
&\simeq
\left(
1-\frac{\Omega_y^2+\Omega_z^2}{2},
\,\Omega_y,\,\Omega_z
\right).
\end{aligned}
\label{Omega_AP}
\end{equation}
and
\begin{equation}
\begin{aligned}
\boldsymbol{\Omega}'
&=
\left(
\sqrt{1-{\Omega'_y}^2-{\Omega'_z}^2},
\,\Omega'_y,\,\Omega'_z
\right)
\\
&\simeq
\left(
1-\frac{{\Omega'_y}^2+{\Omega'_z}^2}{2},
\,\Omega'_y,\,\Omega'_z
\right).
\end{aligned}
\label{Omega7_AP}
\end{equation}
We introduce the differences
\begin{equation}
\Delta\Omega_y=\Omega'_y-\Omega_y,
\qquad
\Delta\Omega_z=\Omega'_z-\Omega_z,
\label{Delta_1AP}
\end{equation}
and
\begin{equation}
(\Delta\Omega)^2
=
(\Delta\Omega_y)^2+(\Delta\Omega_z)^2.
\label{Delta_12AP}
\end{equation}
Using the above expansions, it gives
\begin{equation}
\mathbf{\Omega}\times\mathbf{\Omega}'
\simeq
\left(
\Omega_y\Omega'_z-\Omega_z\Omega'_y,
-\Delta\Omega_z,
\Delta\Omega_y
\right).
\label{Delta_123AP}
\end{equation}
Since the $x$ component is of second order in the transverse components,
the leading-order expression for $\mathbf{e}_3$ is
\begin{equation}
\mathbf{e}_3
\simeq
\frac{1}{\Delta\Omega}
\left(
0,
-\Delta\Omega_z,
\Delta\Omega_y
\right).
\label{Delta_1234AP}
\end{equation}
It follows that
\begin{equation}
(\mathbf{e}_3\cdot\mathbf{e}_x)^2
\simeq 0,
\label{e3xxAP}
\end{equation}
\begin{equation}
(\mathbf{e}_3\cdot\mathbf{e}_y)^2
\simeq
\frac{(\Delta\Omega_z)^2}{(\Delta\Omega)^2},
\label{e3yyAP}
\end{equation}
and
\begin{equation}
(\mathbf{e}_3\cdot\mathbf{e}_z)^2
\simeq
\frac{(\Delta\Omega_y)^2}{(\Delta\Omega)^2}.
\label{e3zzAP}
\end{equation}

The geometrical factors in Eqs.~(\ref{A_G1})--(\ref{A_G3}) for splay, twist, and bend deformations are
given by
\begin{equation}
G_i^{\rm MS}
=
\alpha^4
\left(\frac{L_z}{D}\right)
(\mathbf{\Omega}\cdot\mathbf{e}_y)
(\mathbf{\Omega}'\cdot\mathbf{e}_y)
\left[
1+A_{\rm cyc}
(\mathbf{e}_3\cdot\mathbf{e}_i)^2
\right],
\label{G123_AP}
\end{equation}
where $\mathbf{e}_i=\mathbf{e}_y$, $\mathbf{e}_z$, and
$\mathbf{e}_x$ for $i=1$, $2$, and $3$, respectively. Retaining the
leading-order terms in the transverse components yields
\begin{equation}
G_1^{\rm MS}
\simeq
\alpha^4
\left(\frac{L_z}{D}\right)
\Omega_y\Omega'_y
\left[
1+
A_{\rm cyc}
\frac{(\Delta\Omega_z)^2}{(\Delta\Omega)^2}
\right],
\label{G1_AP}
\end{equation}
\begin{equation}
G_2^{\rm MS}
\simeq
\alpha^4
\left(\frac{L_z}{D}\right)
\Omega_y\Omega'_y
\left[
1+
A_{\rm cyc}
\frac{(\Delta\Omega_y)^2}{(\Delta\Omega)^2}
\right],
\label{G2_AP}
\end{equation}
and
\begin{equation}
G_3^{\rm MS}
\simeq
\alpha^4
\left(\frac{L_z}{D}\right)
\Omega_y\Omega'_y.
\label{G3_AP}
\end{equation}
The strong-alignment expressions for $G_1^{\rm MS}$, $G_2^{\rm MS}$, and $G_3^{\rm MS}$ reveal that the anisotropic corrections to the three geometrical factors involve different components of the relative orientational differences. Specifically, the correction to the splay geometrical factor $G_1^{\rm MS}$ is proportional to $(\Delta\Omega_z)^2/(\Delta\Omega)^2$, whereas that to the twist geometrical factor $G_2^{\rm MS}$ is proportional to $(\Delta\Omega_y)^2/(\Delta\Omega)^2$. In contrast, the bend geometrical factor $G_3^{\rm MS}$ has no anisotropic correction at this order. Thus, the shape anisotropy of the cylindrical interaction volume affects the three deformation modes differently through the transverse components of the relative orientational differences.

The Frank elastic constants are calculated from Eq.~(\ref{A13}):
\begin{equation}
\begin{aligned}
K_{ii}
&=
\frac{c_0}{8}\nu_M\phi^2
\iint
f'(\mathbf{\Omega}\cdot\mathbf{n})
f'(\mathbf{\Omega}'\cdot\mathbf{n})
\\
&\quad\times
P_2(\mathbf{\Omega}\cdot\mathbf{\Omega}')
G_i^{\rm MS}
\,d\mathbf{\Omega}\,d\mathbf{\Omega}'.
\end{aligned}
\label{Kii_AP}
\end{equation}
Therefore, the difference between two elastic constants is controlled by
$A_{\rm cyc}$ and the corresponding orientational average. For example,
\begin{equation}
\begin{aligned}
K_{11}-K_{22}
&\simeq
\mathcal{C}A_{\rm cyc}
\iint
\mathcal{W}(\mathbf{\Omega},\mathbf{\Omega}')
\\
&\quad\times
\frac{
(\Delta\Omega_z)^2-(\Delta\Omega_y)^2
}
{\Delta\Omega^2}
\,d\mathbf{\Omega}\,d\mathbf{\Omega}'.
\end{aligned}
\label{Kii123_AP}
\end{equation}
where
\[
\mathcal{C}
=
\frac{c_0}{8}\nu_M\phi^2
\alpha^4\left(\frac{L_z}{D}\right),
\label{Kii123_C}
\]
and
\begin{equation}
\mathcal{W}(\mathbf{\Omega},\mathbf{\Omega}')
=
f'(\mathbf{\Omega}\cdot\mathbf{n})
f'(\mathbf{\Omega}'\cdot\mathbf{n})
P_2(\mathbf{\Omega}\cdot\mathbf{\Omega}')
\Omega_y\Omega'_y.
\label{Kii123ww_AP}
\end{equation}

At the special aspect ratio
$
\alpha_z={\sqrt{3}}/{2},
$
or 
$
A_{\rm cyc}=0, 
$
the geometrical factors are then identical,
\[
G_1^{\rm MS}
=
G_2^{\rm MS}
=
G_3^{\rm MS},
\]
and we have
$
K_{11}=K_{22}=K_{33}.
$
For $\alpha_z<\sqrt{3}/2$, $A_{\rm cyc}<0$; for $\alpha_z>\sqrt{3}/2$, $A_{\rm cyc}>0$.

For $\alpha_z<\sqrt{3}/2$, the coefficient $A_{\rm cyc}$ is negative and becomes increasingly negative as $\alpha_z$ decreases. Equation~(\ref{G1_AP}) shows that the splay geometrical factor contains an additional contribution proportional to $A_{\rm cyc}(\Delta\Omega_z)^2/(\Delta\Omega)^2$, which is absent in the bend geometrical factor given by Eq.~(\ref{G3_AP}). This negative anisotropic contribution tends to reduce the splay geometrical factor relative to the bend geometrical factor. This tendency is consistent with the numerical result $K_{33}>K_{11}$.
For $A_{\rm cyc}<0$, the twist elastic constant $K_{22}$ changes its relative position from the largest to the smallest among the three Frank elastic constants as the NIT is approached, as indicated by Eq.~(\ref{K123123}). This behavior is qualitatively consistent with the negative anisotropic correction to the factor in $G_2^{\rm MS}$, which is proportional to $A_{\rm cyc}(\Delta\Omega_y)^2/(\Delta\Omega)^2$.

When $\alpha_z>\sqrt{3}/2$, the coefficient $A_{\rm cyc}$ is positive and increases with $\alpha_z$. In this case, the additional contribution to the bracketed factor in the splay geometrical factor (Eq.~(\ref{G1_AP})), proportional to $A_{\rm cyc}(\Delta\Omega_z)^2/(\Delta\Omega)^2$, is positive. This correction tends to enhance the splay geometrical factor relative to the bend geometrical factor as the cylindrical interaction volume is elongated in the $z$-direction. This tendency is consistent with the numerical result $K_{11}>K_{33}$.
For $A_{\rm cyc}>0$, the twist elastic constant $K_{22}$ changes its relative position from the smallest to the largest among the three Frank elastic constants as the NIT is approached, as indicated by Eq.~(\ref{k123_az_big}). This behavior is qualitatively consistent with the positive anisotropic correction to the factor in $G_2^{\rm MS}$, which is proportional to $A_{\rm cyc}(\Delta\Omega_y)^2/(\Delta\Omega)^2$.

%%%%%%%%%%%%%%%%
%%%%%%  References  %%%%%%%%%%
%%%%%%%%%%%%%%%%

%%%%%%%%%

\end{document}